# Influence of symmetry on Sm magnetism studied on $SmIr_2Si_2$ polymorphs

Michal Vališka, Jiří Pospíšil, Jan Prokleška, Martin Diviš, Alexandra Rudajevová, Ilja Turek and Vladimír Sechovský

*Faculty of Mathematics and Physics, Charles University, DCMP, Ke Karlovu 5, CZ-12116 Praha 2, Czech Republic*

## Abstract

Polycrystalline samples of $SmIr_2Si_2$ formed at room temperature both the low temperature phase (LTP) and the metastable high temperature phase (HTP), respectively, depending on the heat treatment. The samples were studied by X-ray powder diffraction, DTA, specific-heat and magnetization measurements with respect to temperature and magnetic field. The first order LTP ↔ HTP polymorphic phase transition has been determined showing the huge temperature hysteresis of 264°C caused by the high energy barrier due to the change of stacking of Sm, Ir and Si basal plane sheets within the transition. Both polymorphs show indications of antiferromagnetic order at low temperatures. The considerably different magnetic phase transitions determined for the LTP and HTP confirm the strong influence of crystal structure symmetry on magnetism in the two polymorphs. The magnetism in $SmIr_2Si_2$ exhibits typical features caused by the specific behavior of $Sm^{3+}$ ion characterized by energy nearness of the ground state and first excited state and crystal field influence. Interpretation of experimental results is corroborated by results of ab initio electronic structure calculations.

## Keywords

# 1 Introduction

The variety of intriguing physical properties makes the rare earth intermetallics subject of an intensive scientific interest. These compounds exhibit various types of nontrivial magnetic order, heavy-fermion behavior and some of them are superconductors. The magnetic properties are associated with well-localized 4$f$ electrons of the rare earth ion. Samarium intermetallics, which found rich practical applications as permanent magnets, deserve special attention owing to unusual features of the $Sm^{3+}$ ion. When analyzing magnetic data measured for numerous Sm compounds the results by rule considerably differ from the expected behavior of the free $Sm^{3+}$ ion when accounting for the ground state multiplet $J = 5/2$ only. The key reason for the observed discrepancies is that the energy intervals between the consecutive $J$ multiplets are relatively narrow, e.g., the first excited multiplet of $J = 7/2$ is lying only about 1500 K above the ground state multiplet[1]. Therefore, the virtual transitions among the multiplets through the off diagonal - $J$ matrix elements of the Hamiltonian may appear and the thermal populations of the excited $J$ levels has considerable influence on magnetic properties[2].

Moreover the spin and orbital component of the localized 4$f$ moment in the $Sm^{3+}$ ion couple antiparallel ($J = L - S$) owing to the spin-orbit interaction, which yields considerable mutual cancellation for the ground multiplet of $J = 5/2$ and a reduced value (0.85$\mu_B$) of the ground state magnetic moment is expected. Consequently, this makes the contribution of the conduction electrons to the total magnetic moment "more visible"[3-6].

Furthermore, the magnetic behavior of the $Sm^{3+}$ ion is significantly modified in solids by the crystal field (CF) and the exchange interactions [7-13] may lead to multiple magnetic phase transitions, etc. It is obvious that the change of CF due to changing symmetry of the Sm ion neighborhood ligand can dramatically modify the physical properties of a particular compound. A natural option to study impact of the change of the CF symmetry is provided by polymorphic materials. Polymorphism is a material property when compound can adopt two or more crystal structures. It means that the nearest ligands surrounding the rare earth ion are re-arranged yielding different CF symmetry while the chemical composition remains unchanged.

Some of the ternary intermetallics with the general composition $MT_2X_2$, where $M$ stands for a rare-earth, actinide or alkaline earth elements, $T$ is one of the transition or post-transition elements and finally $X$ is a $p$ element, exhibit polymorphism. One of the two polymorphs usually stable at room temperature (RT) is called the low temperature phase

(LTP) and crystallizes in the body centered tetragonal $ThCr_2Si_2$ type structure[14, 15] (space group I4/mmm). The high temperature phase (HTP) adopting the primitive tetragonal $CaBe_2Ge_2$-type structure[15] (space group P4/nmm) is stable above a characteristic temperature of several hundred K above RT. The HTP can be also found at RT when the material has been rapidly cooled from high temperatures.

We focused our interest on $SmIr_2Si_2$ when published lattice parameters for both the polymorphs[16] were the only available information. The reason for lack of any other information may be in rather complicated synthesis of samples due to highly evaporating Sm.

## 2 Materials and methods

The polycrystalline $SmIr_2Si_2$ samples have been prepared from the high-purity elements (Ir = 99.99%, Si = 99.9999% and Sm = 99.9%) by two ways: a) melting the stoichiometric amounts of the three elements (with a 1% Sm mass excess to compensate for evaporation losses during melting) in a monoarc furnace under the high-purity argon atmosphere (6N). The melting was repeated three times for achieving well homogeneous samples. The final product of the melting was hard, reflective, silvery white roughly spherical button with 10 mm in diameter and a mass of about 2.5g. b) Much better samples, however, have been obtained by a two-steps process. In the first step we have prepared the binary IrSi compound by arc melting of stoichiometric amount of pure elements. This step led to a significant reduction of the power required for melting of pure Ir (melting point 2447°C). Melting point of the IrSi is about 500°C less than for pure Ir[17]. Then we have pulverized the IrSi sample to fine powder. In the second we weighted pure Sm (no mass excess) and a corresponding amount of the IrSi powder. The Sm piece was completely covered by powder on the crucible. Already a low arc power (no sign of melting) led to a strong reaction of the IrSi powder with the Sm bulk piece with no sign of Sm evaporation. After that we increased the arc power to melt the product and received a fragile silver reflecting metallic button. Finally, we have re-melted it twice. The resulting sample was used for the following studies.

It has been carefully cut into two halves with a fine wire saw to avoid creation of additional internal stresses. One half has been left in the "as-cast" state (expected to be HTP) and the second half has been wrapped in a tantalum foil (99.9%), sealed in a quartz tube and annealed at 900°C for 7 days under the vacuum $1 \cdot 10^{-6}$ mbar and subsequently cooled slowly down to obtain the LTP. Both the samples (as-cast and annealed) were characterized by the

X-ray powder diffraction (XRPD) method. The XRPD patterns have been recorded at room temperature on a Bruker D8 Advance equipped with a monochromatic providing the CuK$\alpha$ radiation. The diffraction patterns were evaluated by the standard Rietveld technique[18] using the FullProf/WinPlotr software[19]. The sample composition has been verified by chemical analysis using the scanning electron microscope (SEM) Tescan Mira I LMH equipped by an energy dispersive X-ray detector (EDX) Bruker AXS[20]. The samples of shapes appropriate for all measurements were cut from the ingot by the fine wire saw.

Differential thermal analysis (DTA) measurements were performed using Setaram SETSYS-2400 CS instrument over the range from room temperature to 1400°C. The heating and cooling rates were set to 5°C/min.

The samples for heat capacity measurement were prepared in the shape of small plates 1.5 x 1.5 x 0.5 mm$^3$. Relaxation method was used for heat capacity measurements. Magnetization measurements were carried out on randomly oriented fixed powders. All measurements were performed using a PPMS (Physical Property Measurement System delivered by Quantum Design) and MPMS (Magnetic Property Measurement System, also Quantum Design) devices[21].

To obtain direct information about the ground state electronic properties, we have carried out first principles theoretical calculations. The ground state electronic structure was calculated on the basis of density functional theory (DFT) within local spin density approximation (LSDA)[22] since LSDA provides the best theoretical volume for YIr$_2$Si$_2$[23]. For this purpose, we used the full potential augmented plane wave plus local orbitals method (APW+lo) as implemented in the latest version of the original WIEN code[24]. The calculations were scalar relativistic and were performed with the following parameters. Non-overlapping atomic spheres (AS) radii of 2.8, 2.3 and 1.6 a.u. (1.u. = 52.9177 pm) were taken for Sm, Ir and Si, respectively. The basis for expansion of the valence states (less than 8 Ry below Fermi energy) consisted of more than 1300 basis function (more than 260 APW per atom) plus Sm (5$s$, 5$p$), Ir (5$s$, 5$p$, 4$f$) and Si 2$p$ local orbitals. The 4$f$ electrons are localized; therefore, the Sm 4$f$ states were treated in the open core approximation with the stable atomic configuration 4$f^5$. The wave functions in the AS region were expanded up to $l = 12$ and for the interstitial charge density GMAX = 14 was used. The Brillouin zone (BZ) integrations were performed with the tetrahedron method[24] on a 163 (I4mmm, 2000 k-points in the full BZ) and 180 special k-point mesh (P4/mmm, 2000 k-points in the full BZ). We carefully tested the convergence of the presented results with respect to the parameters mentioned and found them to be fully sufficient for all presented characteristic for both phases of SmIr$_2$Si$_2$. First principles calculations of the CF interaction were performed using the method described in Ref.[25]. Within this method, the electronic structure and the corresponding distribution of the ground state charge density are obtained using the full potential APW+lo method. The CF parameters originate from the aspherical part of the total single particle DFT potential in the crystal. To eliminate the self-interaction, the self-consistent procedure is performed with the 4$f$ electrons in the core[25], which is the open core approximation used in this work.

## 3 Results and Discussion

### 3.1 Crystal structure analysis

Both the "as-cast" and "annealed" samples have been characterized by the X-ray powder diffraction. The XRPD data refinement revealed that the annealed sample consisted of only the LTP whereas the as-cast sample contains majority of the HTP with some volume fraction (10 – 30%) of the LTP. The latter result indicates that, similar to the case of NdIr$_2$Si$_2$ reported by Mihalik et al.[26], cooling of the sample after the arc melting was not fast enough to avoid formation of LTP. After few trials we have observed that reducing the mass of the melt to 0.5 g allows much faster cooling of the sample so that the as single HTP sample is obtained. The X-ray powder patterns of the single phase as-cast (HTP) and annealed (LTP) samples are displayed in Figure 1. Only the single-phase samples have been used for further measurements.

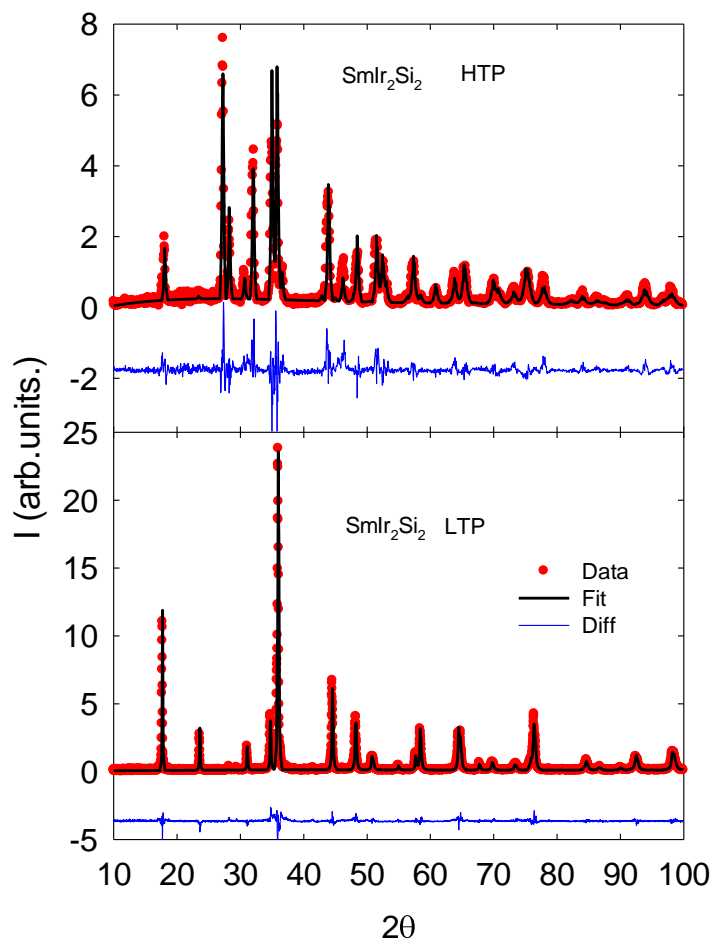

Figure 1. XRPD patterns of the SmIr$_2$Si$_2$ compound. The upper panel represents the pattern of the as-cast sample – HTP phase. The lower panel represents the pattern of the annealed sample – LTP phase. All phases are well described by used model.

The further detailed analysis of the diffraction pattern for the "as-cast" sample revealed the mixture between the Ir and Si atoms in the *2c* positions, which, however, amounts only of 2 %. The LTP model for the "annealed" sample describes perfectly measured pattern. There remained no unlabelled reflections when describing the data with LTP. We can suppose that the sample successfully transformed to the LTP with no trace of the HTP or mixture between the Ir and Si atoms. The obtained atomic coordinates and the lattice parameters are digestedly listed in Table 1 and Table 2. Structural models used in our work well correspond with prior crystallographic data[16].

Table 1. Crystallographic parameters of the HTP of SmIr$_2$Si$_2$.

| | SmIr$_2$Si$_2$ – HT | | a (Å) | 4.1216(5) |
|---|---|---|---|---|
| | CaBa$_2$Ge$_2$ structure type | | c (Å) | 9.8235(3) |
| Atoms | Symmetry | x | y | z |
| Ir | 2c | ¼ | ¼ | 0.1318 |
| Si | 2c | ¼ | ¼ | 0.4227 |
| Sm | 2c | ¼ | ¼ | 0.7521 |
| Ir | 2b | ¾ | ¼ | ½ |
| Si | 2a | ¾ | ¼ | 0 |

Table 2. Crystal structure parameters of the LTP of SmIr$_2$Si$_2$.

| | SmIr$_2$Si$_2$ – LT | | a (Å) | 4.0671(2) |
|---|---|---|---|---|
| | ThCr$_2$Si$_2$ structure type | | c (Å) | 10.0182(7) |
| Atoms | Symmetry | x | y | z |
| Si | 4e | 0 | 0 | 0.37388 |
| Ir | 4d | 0 | ½ | ¼ |
| Sm | 2a | 0 | 0 | 0 |

The scanning electron microscope equipped by SE and BSE detectors[20] and EDX analyzer were used to confirm composition of the samples. Different intensities on the surface in BSE pictures obtained by SEM revealed that polycrystalline sample (as cast) consists of two phases with different composition. The elementary analysis confirmed the majority of the SmIr$_2$Si$_2$ phase and tiny amount of a phase 3 - 5 volume % richer on Ir.

We have used differential thermal analysis (DTA) for investigation of the phase transitions between the HTP and LTP. Figure 2 shows the DTA curve for the "as-cast" sample which undergone one thermal cycle from room temperature (RT) to 1400°C and back to RT. Two peaks can be recognized on the heating branch. First one at the temperature $T_1$ (414°C) corresponds to the thermal relaxation from the metastable (quenched) HTP to LTP (similar as reported by Mihalik et al.[27] for PrIr$_2$Si$_2$). The second anomaly at $T_2$ corresponds to the transition from LTP to the HTP, which is stable above 1194°C. If we follow the cooling branch we can recognize that the sample transforms from the HTP to the LTP at $T_3$ (930°C).

The DTA curve for the annealed sample has the same character except for the missing $T_1$ anomaly because the sample already consists at the beginning of the cycle at RT of the LTP phase. The phase change on the heating branch is only the LTP → HTP transition at $T_2$. The same holds for the second thermal cycle for the "as cast" sample, which after the first thermal cycle also consists of the LPT phase only.

Figure 2. The DTA curve recorded for the first thermal cycle of the "as cast" sample, the arrows show the course of temperature change.

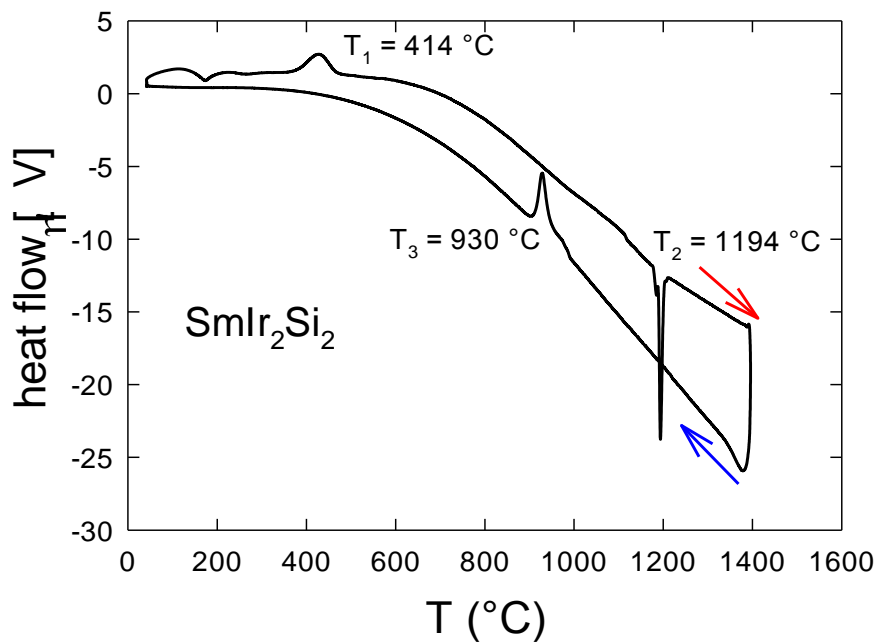

The DTA analysis also shows that the temperature interval 800 – 900°C is the most suitable for the annealing process of the $SmIr_2Si_2$. This temperature region is high enough to enable the transformation from the metastable HTP to the LTP but at the same time it is low enough ($<T_2$) to avoid the reverse transformation from the LTP to the HTP. It should be emphasized the LTP ↔ HTP transition in $SmIr_2Si_2$ has large hysteresis of 264°C (= $T_2 - T_3$) which corroborates that we deal with the first order transition. It has been mentioned that both phases are the variants of one mother structure (the $BaAl_4$ - type tetragonal structure). They are formed by *RE*, *T* and *X* basal-plane atomic layers stacked along the c-axis. Different sequence of these layers differentiates the two phases. We observe *RE-X-T-X-RE-X-T-X-RE*... for the $ThCr_2Si_2$- type (LTP) and *RE-T-X-T-RE-X-T-X-RE*... for the $CaBe_2Ge_2$- type (HTP). Therefore in the LTP the *RE* and *T* atoms are separated by the *X* atom layers, whereas the *T* and *X* atom layers are interchanged in the second half of the unit cell in the HTP. It can be expected that the crystal phase transition between the two structures costs a lot of energy and consequently a large temperature hysteresis can be expected. R. Hoffmann and C. Zheng[28] performed calculations of the *X-X* bonds in the $AB_2X_2$ compounds and they concluded that there exists energy minimum for both the *X-X* bonds present (bond state) and *X-X* bonds broken (anti-bonding state). Thus there may be a structural phase transition for some specific *A*, *B* and *X* element configuration. Our work experimentally confirms this hypothesis. The theoretical existence of a double-well-like ground state explains the presence of the exactly and only two different crystallographic structures separated by a sharp phase transition. Furthermore, based on the mentioned model of Hoffman and Zheng, we can also propose a qualitative explanation of the observed large temperature hysteresis in this compound. At low temperatures the energy minimum for the Si-Si short distance along the c-axis (a bond state in the LTP, typically 2.3 Å long) occurs at lower energy than the minimum for the corresponding long distance (a no-bond state in the HTP, over 3.9 Å long). Therefore, while the temperature is increased, the energy minimum for the no-bond state is lowered. However, due to the energy barrier between the two states, the compound persists in the LTP structure (bond state). By further temperature increase, the bond state energy minimum gets shallower and the no-bond state energy minimum gets deeper, consequently the barrier between the two minima vanishes and the compound transforms into the HTP structure. Taking into account identical construction for the cooling back to room temperature, we end up with the hysteretic behavior.

## 3.2 Specific heat data analysis

Both the "as-cast" and "annealed" samples have been subjects of the heat capacity measurements performed in wide temperature range (0.35 - 300 K) and with applied various external magnetic fields up to 14 T. These measurements confirmed the strong crystal-field influence on magnetism in Sm compounds. In the case of the HTP we have observed one magnetic transition, namely at 22 K. The LTP exhibits rather complicated behavior with three magnetic phase transitions at critical temperatures $T_{C3}$ = 1.9 K, $T_{C2}$ = 6.2 K and $T_{C1}$ = 38 K, as can be seen in Figure 3. In addition one can see a small hint of a transition below temperature 0.5 K in both Sm polymorphs - see the inset in Figure 3. It can denote another magnetic phase transition or more likely the nuclear Schottky contribution from Ir to the specific heat.

Application of an external magnetic field up to 14 T does not shift any of the observed transitions except for a weak shift of transition from $T_{C2}$ = 6.2 K to $T_{C2}$ = 5.6 K at 14 T in the LTP (Figure 4). The magnetic transition in HTP polymorph is untouched up to magnetic field 10 T.

Figure 3. Temperature dependence of the heat capacity of the $SmIr_2Si_2$ polymorphs, LTP and HPT. The transitions are marked by arrows.

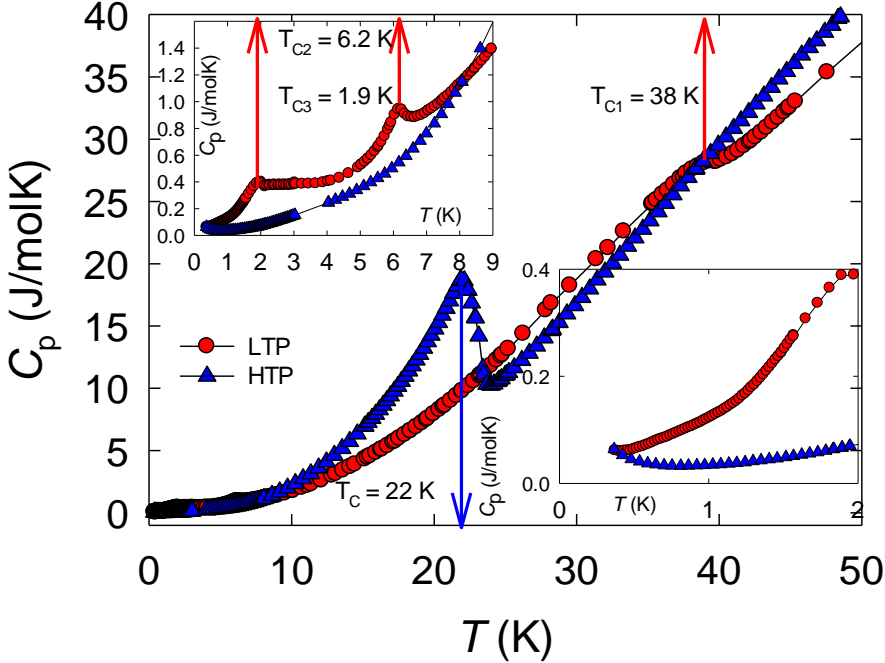

Figure 4. The effect of an applied magnetic field on the magnetic transition in both polymorphs.

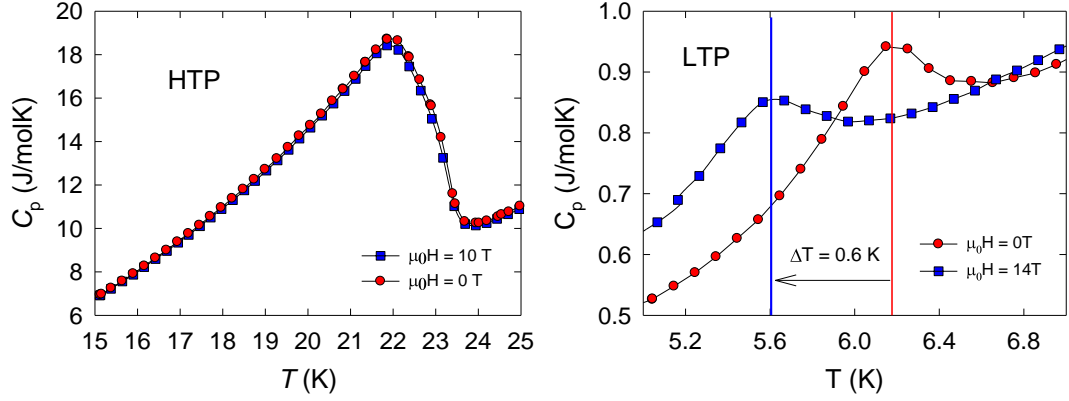

The measured specific heat has been considered as the sum of contributions (Eq.1):

$$C_p = C_{el} + C_{ph} + C_{mag} + C_{sch} \qquad (Eq.1)$$

electronic $C_{el}$, phonon $C_{ph}$, magnetic $C_{mag}$ and Shottky $C_{sch}$ contribution, respectively.

The most significant is the phonon contribution mainly at high temperatures and this part has been evaluated within the Debye and simplified isotropic Einstein model (Eq. 2). Results are listed in the Table 3 and 4.

$$C_{ph} = R\left( \frac{1}{1-\alpha_D T} C_D + \sum_{i=1}^{3N-3} \frac{1}{1-\alpha_E T} C_E \right) \qquad (Eq.2)$$

| branches | degeneracy | $\Theta$ [K] | $\alpha$ [$10^{-4}$ K$^{-1}$] |
|---|---|---|---|
| $\theta_D$ |  | 255 | 4.0 |
| $\theta_{E1}$ | 4 | 125 | 4.0 |
| $\theta_{E2}$ | 4 | 270 | 4.0 |
| $\theta_{E3}$ | 4 | 550 | 4.0 |

Table 3. The values and degeneracy of branches of the phonon contribution to the total specific heat of the LTP evaluated on the basis of the simplified isotropic Einstein and Debye models.

| branches | degeneracy | $\Theta$ [K] | $\alpha$ [$10^{-4}$ K$^{-1}$] |
|---|---|---|---|
| $\theta_D$ |  | 265 | 5.0 |
| $\theta_{E1}$ | 4 | 115 | 4.0 |
| $\theta_{E2}$ | 4 | 230 | 3.0 |
| $\theta_{E3}$ | 4 | 780 | 2.5 |

Table 4. The values and degeneracy of branches of the phonon contribution to the total specific heat of the HTP evaluated on the basis of simplified isotropic Einstein and Debye models.

In reality, the Einstein parts of phonon spectrum (acoustic branches) of both polymorphs are anisotropic and their energy and degeneracy is a characteristic parameter for every unique point in reciprocal space. Nevertheless the used simplified isotropic model gives reasonable results to subtract phonon part from the raw specific-heat.

$$C_e = \frac{2nk_B^2 T}{E_F} = \gamma T \quad (Eq.3)$$

The electronic contributions to the total specific heat (Eq.3) (Sommerfeld coefficient gamma) has been found $\gamma = 10$ mJ/mol K$^2$ for both LTP and HTP, respectively. The $\gamma$-values are typical for Sm intermetallics compounds [13, 29, 30].

$$C_{sch} = \frac{R}{T^2} \left( \frac{\sum_{i=0}^{n} \Delta_i^2 \exp\left[-\frac{\Delta_i}{T}\right]}{\sum_{i=0}^{n} \exp\left[-\frac{\Delta_i}{T}\right]} - \left( \frac{\sum_{i=0}^{n} \Delta_i \exp\left[-\frac{\Delta_i}{T}\right]}{\sum_{i=0}^{n} \exp\left[-\frac{\Delta_i}{T}\right]} \right)^2 \right) \quad (Eq.4)$$

The magnetic entropy associated with the magnetic ordering is considerably smaller than $R\ln 2$ in low temperatures, where R is the gas constant, which is usually expected for the doublet ground state of the Sm$^{3+}$ ion. The value of $R\ln 2$ expected for a ground state doublet is reached at around 25 K for HTP and 40 K for LTP, respectively. Above these temperatures the magnetic entropy still strongly increases and shows a gradual tendency to saturation at the value of $R\ln 6$ expected for the full population of the Sm$^{3+}$ multiplet split by crystal field.

As a next step we have evaluated the Schottky contribution (Eq.4). The ground state is according to our results split by the CF on the three Kramer's doublets for both compounds as is presented in the Table 5.

| levels | energy [K] | energy [K] |
| --- | --- | --- |
|  | LTP | HTP |
| $\Delta_1$ | 0 | 0 |
| $\Delta_2$ | 40 ± 15 | 120 ± 20 |
| $\Delta_3$ | 115 ± 30 | 250 ± 40 |

Table 5. Energy of the 3 Kramer's doublets after CF splitting of the degenerated ground state.

A huge energy gap of 120 K has been observed between the ground and the first excited state in the case of HTP polymorph, which is in a good agreement with findings in [9] and our theoretical calculations discussed later.

### 3.3 Magnetization measurements

The magnetization was measured as function of temperature and magnetic field for the "as-cast" and annealed sample, respectively (see Figure 5 and 6). In both cases the magnetization at 2 K is very weak, reaching in the maximum field of 7 T only ≈ 0.035 μ$_B$/f.u. In fields above 2.5 T the magnetization varies almost linearly with magnetic field and with no tendency to saturation. The M vs. B dependence is strongly nonlinear in low fields. For the HTP phase has been recorded hysteresis closing in fields above 2.5 T. For the LTP phase the hysteresis is negligible. The maximum measured magnetic moment in both polymorphs in 7 T is apparently more than an order lower than the expected value of the saturated magnetic moment of 0.71 μ$_B$/f.u for free Sm$^{3+}$ ion. Presuming that the specific-heat anomalies at 22 K and 38.9 K, respectively, reported above for the HTP and LTP phase, respectively, represent magnetic phase transitions in an antiferromagnetic ordering one expects observing some metamagnetic transitions on the low temperature magnetization curves. Since we have observed no sign of field induced magnetic transitions occurred in our magnetization data so we presume that much stronger magnetic field is needed for reaching the saturated value of the magnetic moment. Our data reflect typical behavior for many Sm intermetallics, e.g. like SmCu$_2$[13, 31], which is usually conceived in terms of strong influence of the crystal field and mixing of the low lying excited multiplet $J = 7/2$ with the ground multiplet $J = 5/2$.

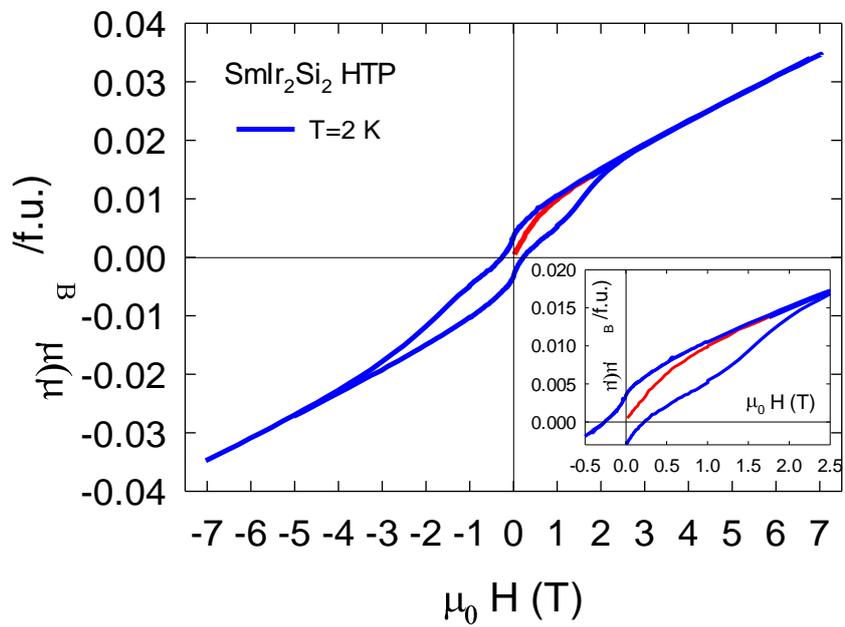

Figure 5. Hysteresis loops of the SmIr$_2$Si$_2$ HTP sample. The insets show low magnetic field detail.

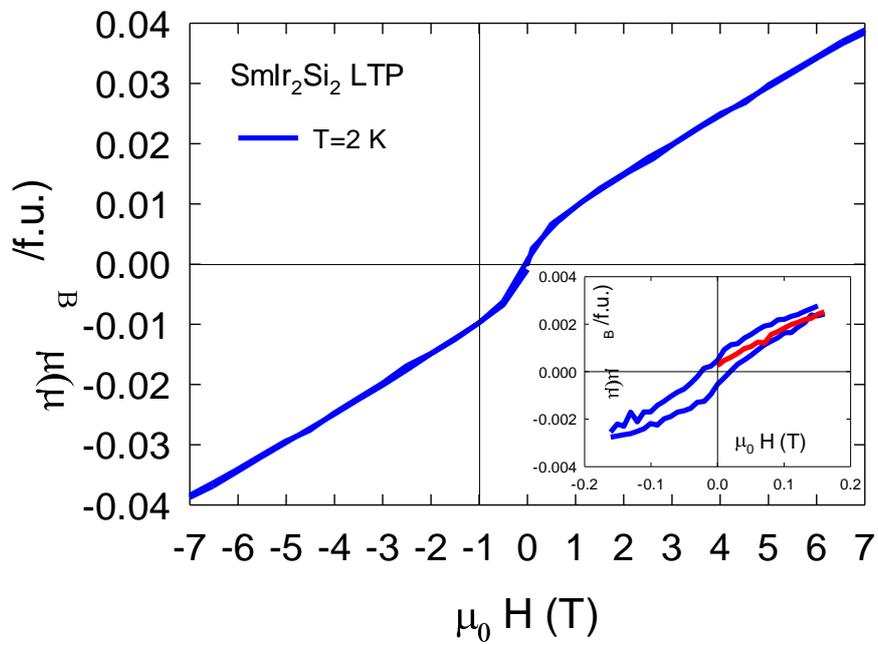

Figure 6. Hysteresis loops of the SmIr$_2$Si$_2$ LTP sample. The insets show low magnetic field detail.

The temperature dependence of the inverse susceptibility shown in Figure 7 is strongly nonlinear for both polymorphs and can be formally well fitted by the modified Curie-Weiss law (MCW)[13, 31] (Eq.5).

$$\chi = \chi_0 + \frac{C}{T - \theta_P} \quad \text{(Eq. 5)}$$

in the temperature range 70 – 230 K for the HTP and 70 – 200 K for the LTP, where the temperature-independent term $\chi_0$ may be attributed to the temperature independent van Vleck contribution which is due to the small energy difference between the ground-state multiplet $J = 5/2$ and the first excited multiplet $J = 7/2$. In the case of the HTP the fit yields the values of the paramagnetic Curie temperature $\theta_p$ = -52.3 K and the effective Sm moment $\mu_{eff}$ = 0.85 $\mu_B$, which is in a good agreement with the value of 0.85 $\mu_B$ expected for the $Sm^{3+}$ free ion and $\chi_0$ = 4.17 * $10^{-9}$ $m^3mol^{-1}$. The considerable deviation from the fit at higher temperatures can be attributed to population of the higher excited multiplet.

Similar susceptibility behavior has been found for the LT-polymorph. The fit yields the values for the paramagnetic Curie temperature $\theta_p$ = -51.7 K, the effective Sm moment $\mu_{eff}$ = 0.87 $\mu_B$, and $\chi_0$ = 2.37 * $10^{-8}$ $m^3mol^{-1}$.

Unfortunately, magnetization data on polycrystalline samples cannot bring sufficient information regarding the magnetocrystalline anisotropy or magnetic structures of the individual magnetic phases. We are also aware of the fact that the fitted MCW parameters may be far different from parameters which might be determined from fitting the intrinsic anisotropic susceptibility data measured on a single crystal if available.

Figure 7. The temperature dependence of the inverse susceptibility for the HTP (left panel) and for the HTP (right panel) of the $SmIr_2Si_2$.

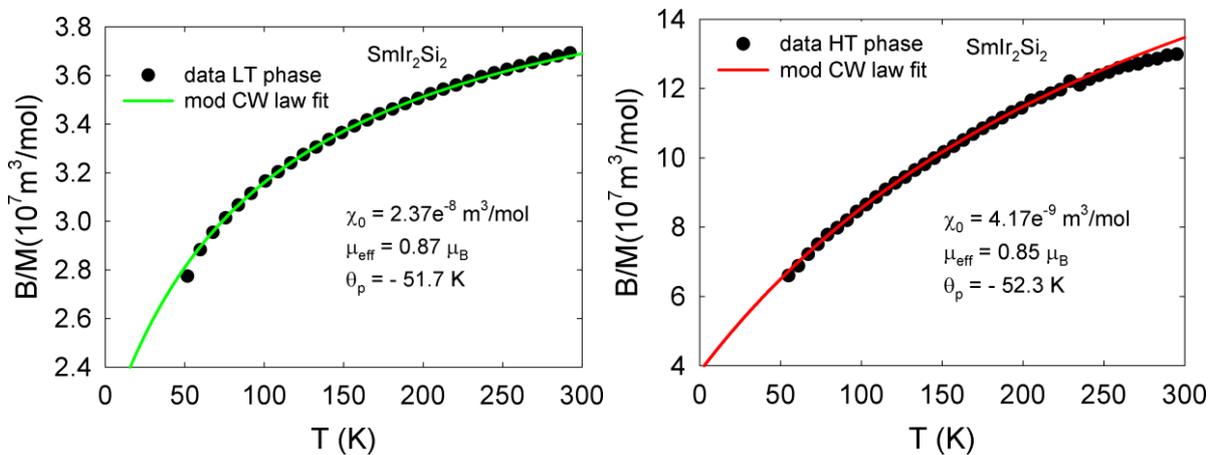

Let us now turn to first principles calculations. The first principles CF calculations provide CF parameters summarized in Table 6. Using these parameters we have found that the HTP and LTP have completely different magnetic properties. For instance the magnetic susceptibility of HTP is higher than LTP (see Figure 8). There are also crossings of susceptibility for both LTP and HTP around 120 K. The most interesting feature is that the easy magnetization axis is *c* in the case of the LTP while the a-axis is the easy magnetization axis for the HTP (Figure 8). We have also calculated the values of polycrystalline reciprocal susceptibility for both polymorphs (Figure 9) with reasonable agreement with experimental data (Figure 7). It is unique influence of the symmetry on such crucial magnetic parameter. Therefore the experimental study of $SmIr_2Si_2$ single crystals is highly desirable.

| Kelvin | I4/mmm | P4/nmm |
|---|---|---|
| $A_2^0$ | 64 | -329 |
| $A_4^0$ | -24.2 | -0.27 |
| $A_4^4$ | 471 | -268 |
| $A_6^0$ | 1.46 | -0.06 |
| $A_6^4$ | -38.5 | -43.8 |

Table 6. CF parameters for the LTP and HTP calculated from first principles.

We have also calculated the splitting of the ground state multiplet by CF. We have found that the degenerated ground state is split to three Cramer's doublets in both compounds however with entirely energy spectra. The energies of the separated levels are summarized in Table 7. The calculated spectrum of the HTP polymorph is characterized by wide gap of 170 K between the ground state and the first excited doublet. The calculated data are in reasonable agreement with our experimental specific heat data.

| Levels | energy [K] | energy [K] |
|---|---|---|
|  | LTP | HTP |
| $\Delta_1$ | 0 | 0 |
| $\Delta_2$ | 10 | 170 |
| $\Delta_3$ | 80 | 250 |

Table 7. Energy of the 3 Kramer's doublets after CF splitting of the degenerated ground state for both compounds.

Figure 8. Magnetic reciprocal susceptibility calculated for HTP and LTP. The full (dashed) lines correspond to the calculated data for the c-axis (a-axis).

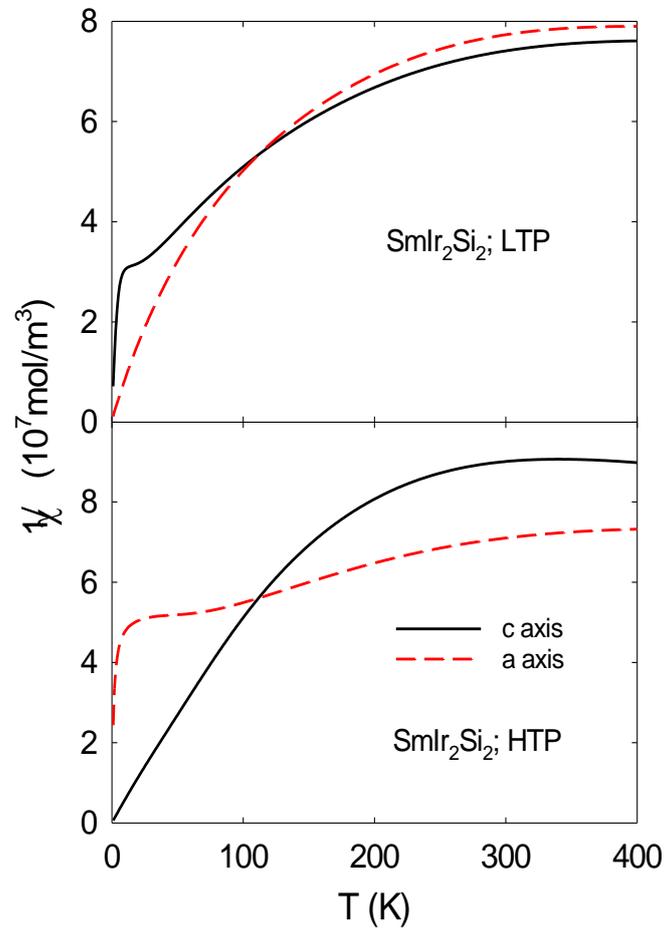

Figure 9. Polycrystalline reciprocal magnetic susceptibility calculated for HTP and LTP. The full (dashed) lines correspond to the calculated data for the HTP (LTP).

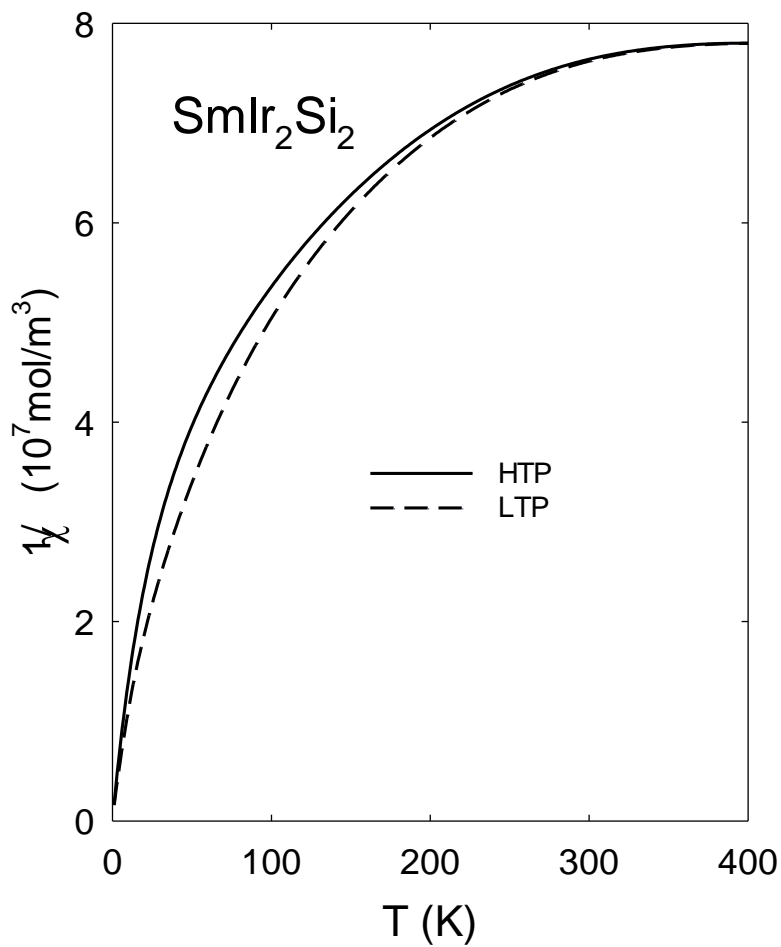

We have also found difference in the electronic structure between the LTP and HTP when the energy gap in the energy interval (-8) – (-7) eV was found in the LTP (Figure 10 and 11).

Figure 10. Total DOS and atom-projected DOS of HTP SmIr$_2$Si$_2$. The projected Sm DOS, both Ir, both Si and the interstitial region (dashed line) are shown. Fermi level is put at zero energy.

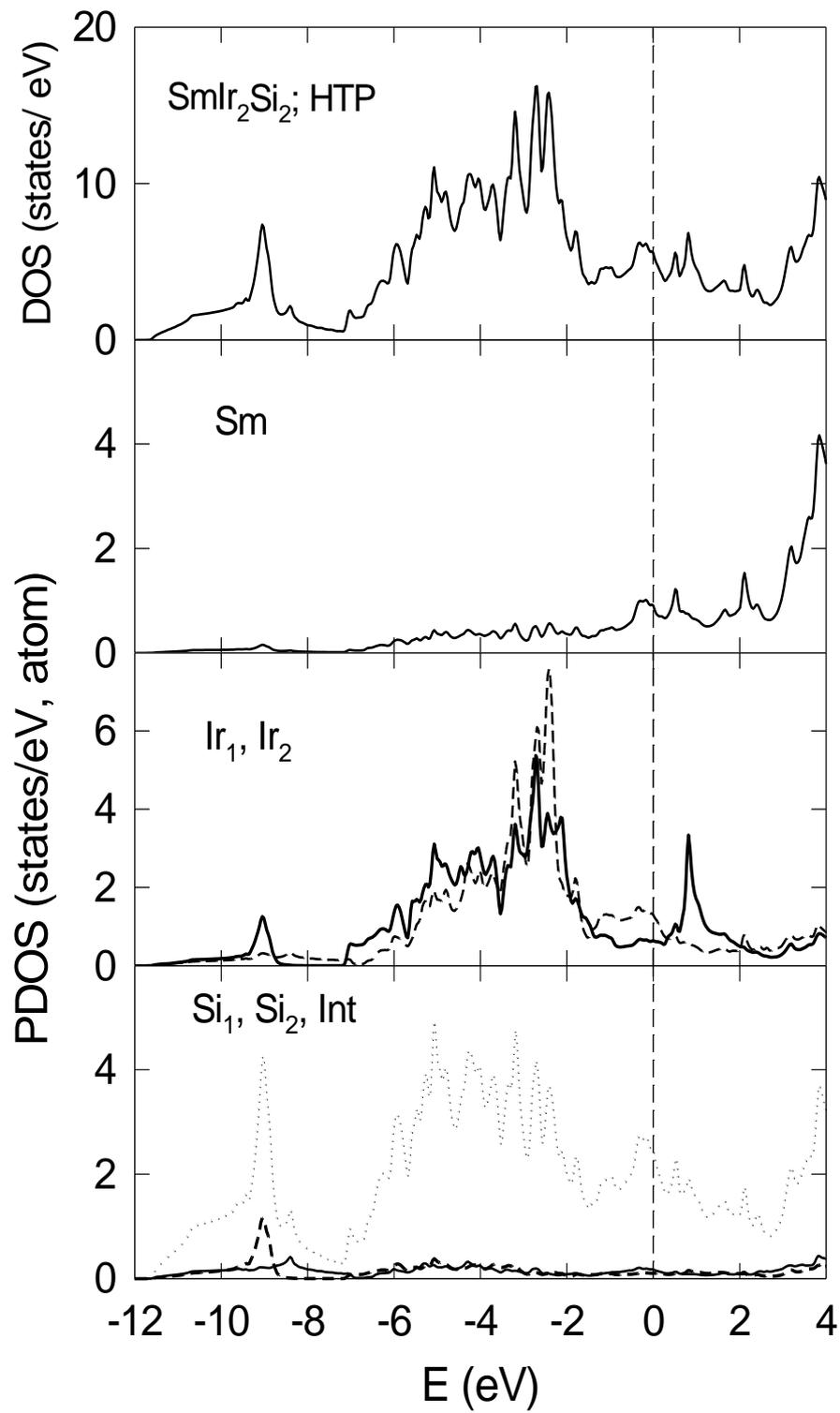

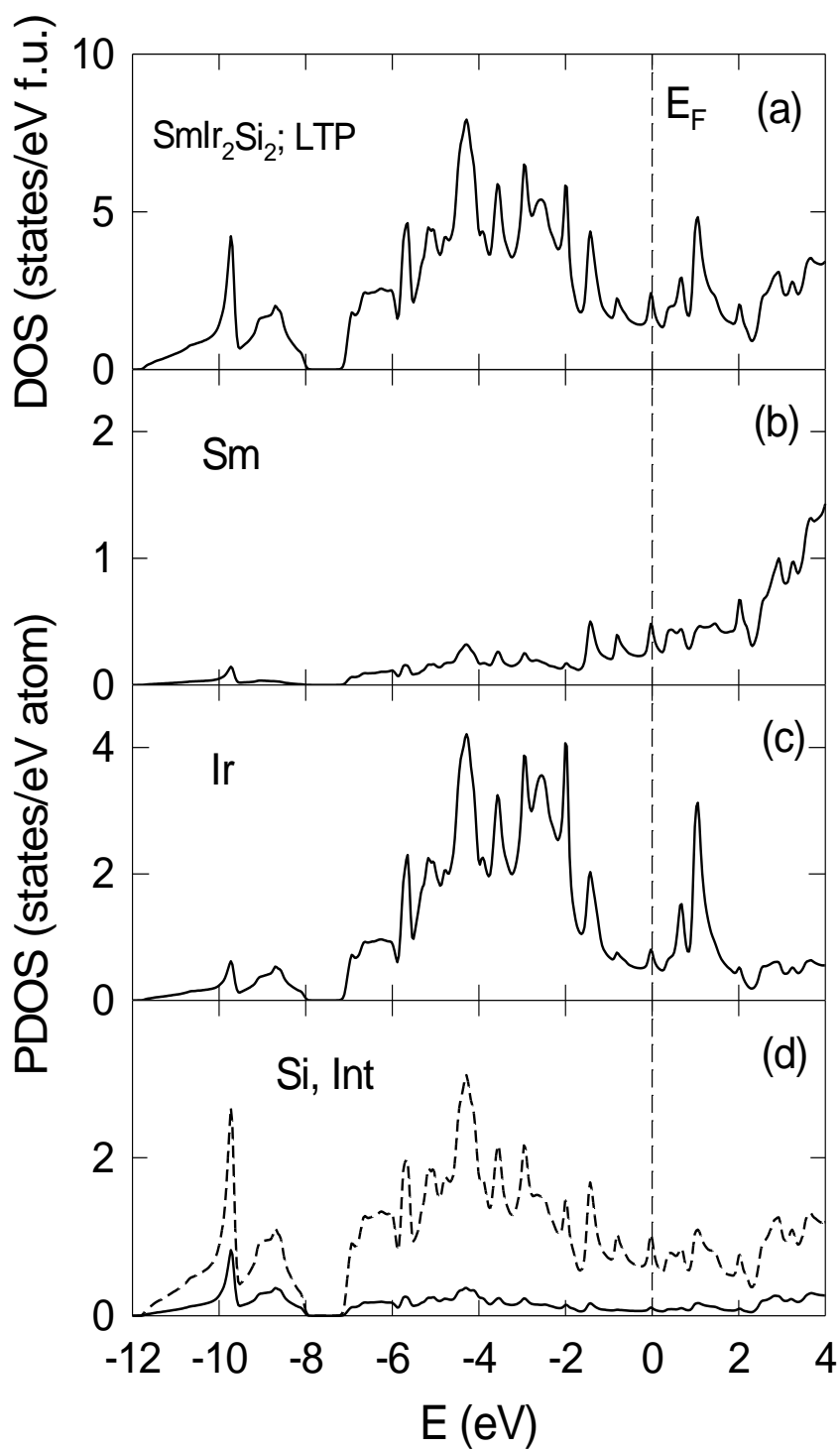

Figure 11. Total DOS and atom-projected DOS of LTP SmIr$_2$Si$_2$. The projected Sm DOS, Ir, Si and the interstitial region (dashed line) are shown. Fermi level is put at zero energy.

# Conclusions

We have successfully synthesized the SmIr$_2$Si$_2$ compound in polycrystalline form in both structure polymorphs. We have studied the thermal treatment process in detail and established the characteristic structural transformation temperatures between the LTP and HTP, which are crucial for the annealing process and preparation of the single phase samples. The first order LPT ↔ HTP phase transition has been determined showing the huge temperature hysteresis of 264°C caused by the high energy barrier due to the change of stacking of Sm, Ir and Si basal plane sheets within the transition.

Both polymorphs show indications of antiferromagnetic ordering below $T_N$ = 22 K and 38.9 K for the HTP and LTP, respectively. The LTP undergoes order-to-order transitions at 6.1 K and 1.9 K, respectively. In addition a small hint of a transition has been indicated below temperature 0.5 K in both polymorphs most likely due to the Ir nuclear contribution. The phase transitions (except for the 6.1-K one) have been found almost intact in magnetic fields up to 10 T. Despite of the limited experimental information obtained on polycrystals we have obtained interesting prediction of the magnetic features of the both phases on the basis of electronic structure calculations like crossing of the easy axis magnetization in each phase and also opposite easy magnetization axis between the LTP and HTP. SmIr$_2$Si$_2$ represents an example of the Sm magnetism with strong influence of crystal symmetry as a critical parameter.

# Acknowledgements

This work was supported by the Czech Science Foundation (Project GACR 202/09/1027) and Charles University grant UNCE 11. Experiments were performed in MLTL (http://mltl.eu/), which is supported within the program of Czech Research Infrastructures (project no. LM2011025).